# PSF and field of view characteristics of imaging and nulling interferometers

François Hénault

UMR 6525 CNRS H. Fizeau – UNS, OCA, Avenue Nicolas Copernic, 06130 Grasse – France

**ABSTRACT**

In this communication are presented some complements to a recent paper entitled "Simple Fourier optics formalism for high angular resolution systems and nulling interferometry" [1], dealing with imaging and nulling capacities of a few types of multi-aperture optical systems. Herein the characteristics of such systems in terms of Point Spread Function (PSF) and Field of View (FoV) are derived from simple analytical expressions that are further evaluated numerically for various configurations. We consider successively the general cases of Fizeau and Michelson interferometers, and those of a monolithic pupil, nulling telescope, of a nulling, Sheared-Pupil Telescope (SPT), and of a sparse aperture, Axially Combined Interferometer (ACI). The analytical formalism also allows establishing the exact Object-Image relationships applicable to nulling PSTs or ACIs that are planned for future space missions searching for habitable extra-solar planets.

## 1  INTRODUCTION

Since the historical writings of Fizeau [2], Stéphan [3] and Michelson [4], multi-aperture optical systems and their imaging properties have been the subject of extensive literature, leading among other to the currently admitted distinction between "Fizeau" and "Michelson" types of interferometer. We use to consider today that the major difference between both concepts relies on the fact that the former obeys a "golden rule" stating that the output pupil of the interferometer must be a scaled replica of its entrance pupil [5-6], while the latter does not. Some new multi-aperture concepts, however, have emerged during the three last decades, such as spaceborne, infrared nulling interferometers or telescopes dedicated to the search of extra-solar planets [7-8] or visible hypertelescopes having unsurpassed imaging capacities [9]. In a recent paper [1] was described a simple first-order, Fourier optics formalism allowing to derive the basic Object-Image relationships of the here above systems as convolution products suitable for fast and accurate calculation. The purpose of this communication is to complete the previous paper, summarizing and simplifying again the formalism in section 2, and providing additional examples in section 3: herein are in turn considered the general cases of Fizeau and Michelson interferometers, of a monolithic, nulling Sheared-Pupil Telescope (SPT), and of a sparse aperture, Axially Combined Interferometer (ACI). Moreover, some imaging properties of the nulling SPTs or ACIs envisaged for future space missions searching for habitable extra-solar planets are further addressed in section 4.

## 2  BASIC RELATIONSHIPS

Here the formalism described in ref. [1] is briefly summarized (and also slightly simplified) in section 2.1, before deriving some theoretical relationships applicable to the Point Spread Function (PSF) and maximal achievable Field of View (FoV) of general multi-aperture optical systems (sections 2.2 and 2.3).

## 2.1 General formalism

Let us consider an optical system designed either for high-angular resolution imaging or nulling interferometry purpose, being composed of N collecting apertures and N recombining apertures. The four attached coordinate systems are depicted in Figure 1, although only three of them will be used herein: an on-sky angular coordinates system (U,V), the entrance pupil reference frame (O,X,Y,Z) where OZ is the main optical axis, and the exit pupil reference frame (O',X',Y',Z) . Let us further assume that:

1) For all indices n comprised between 1 and N, the $n^{th}$ collecting aperture of center $P_n$ is optically conjugated with its associated recombining aperture of center $P'_n$ without any pupilar aberration.

2) All collecting apertures have an identical diameter D and consequently all recombining apertures share the same diameter D'. Practically, it means that all collecting telescopes and optical trains conveying the beams from entrance to exit apertures are identical, which is most often the case for the presently built interferometric facilities, either of Fizeau or Michelson types (see the generic optical layout of Figure 2).

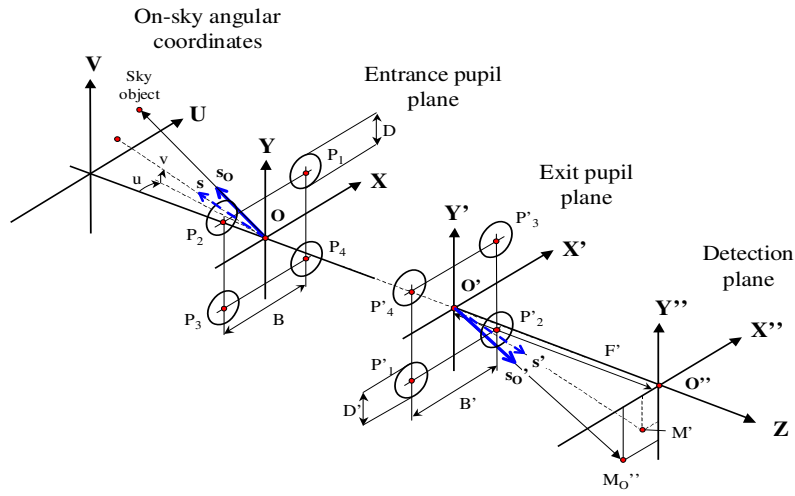

Figure 1: Used reference frames on-sky (U,V), on the entrance pupil plane (O,X,Y) and on the exit pupil plane (O',X',Y'). The coordinate system (O",X",Y") attached to the image plane is not used in this paper.

Let us finally define the following parameters (bold characters denoting vectors):

- **s**  A unit vector of direction cosines ≈ (1,u,v) directed at any point in the sky (corresponding to any point M" in the image plane), where angular coordinates u and v are considered as first-order quantities.
- **$s_O$**  A unit vector of direction cosines ≈ (1,$u_O$,$v_O$) pointed at a given sky object (or an elementary angular area of it)
- O($s_O$)  The angular brightness distribution of an extended sky object
- $\Omega_O$, $d\Omega_O$  The total observed FoV in terms of solid angle, and its differentiating element
- $PSF_T(s)$  The PSF of an individual collecting telescope, being projected back on-sky. For an unobstructed pupil of diameter D, it would be equal to $|2J_1(\rho)/\rho|^2$, where $\rho = kD\|s\|/2$ and $J_1$ is the type-J Bessel function at the first order
- k  The wave number $2\pi/\lambda$ of the electro-magnetic field, assumed to be monochromatic
- λ  The wavelength of the electro-magnetic field
- $a_n$  The amplitude transmission factor of the $n^{th}$ interferometer arm
- $\varphi_n$  A phase-shift introduced along the $n^{th}$ interferometer arm for Optical Path Differences (OPD) compensation or nulling purpose
- m  The optical compression factor of the system, equal to $m = D'/D = F_C/F$ where F and $F_C$ respectively are the focal length of the collecting telescope and of the relay optics (see Figure 2)

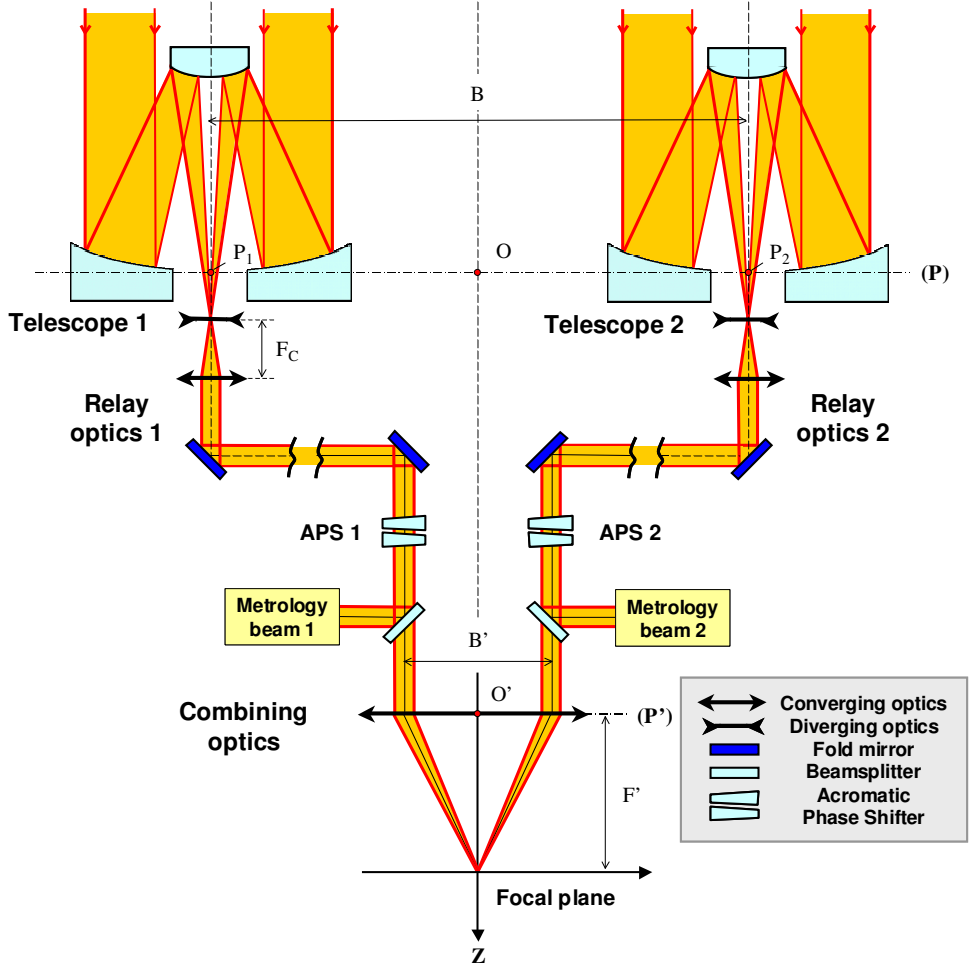

Figure 2: Schematic optical layout of a multi-aperture, high angular resolution interferometer.

Hence according to Ref. [1] the expression of the image I(**s**) formed by the multi-aperture optical system and projected back on-sky writes in a first-order approximation:

$$I(\mathbf{s}) = \iint_{\mathbf{s_O} \in \Omega_O} O(\mathbf{s_O})\, PSF_T(\mathbf{s} - \mathbf{s_O}) \left| \sum_{n=1}^{N} a_n \exp[i\varphi_n] \exp[ik(\mathbf{s_O\,OP_n} - \mathbf{s\,O'P'_n}/m)] \right|^2 d\Omega_O . \qquad (1)$$

This very general Object-Image relationship can only be reduced to convolution products if certain conditions are fulfilled, which are extensively discussed in Ref. [1]. Herein the following sections only deal with other consequences on the PSFs and effective Field of View accessible by the whole system.

## 2.2 Point Spread Functions

We can define the "generalized PSF" of the optical system as simply being the weighting factor of the object brightness function O(**s_O**) under the bi-dimensional integral of Eq. (1) that is:

$$PSF_G(\mathbf{s},\mathbf{s_O}) = PSF_T(\mathbf{s} - \mathbf{s_O}) \left| \sum_{n=1}^{N} a_n \exp[i\varphi_n] \exp[ik(\mathbf{s_O\,OP_n} - \mathbf{s\,O'P'_n}/m)] \right|^2 , \qquad (2)$$

and $\text{PSF}_G(\mathbf{s},\mathbf{s_O})$ presents the particularity of constantly varying with the angular location $\mathbf{s_O}$ of the sky object in the instrument FoV, hence differing significantly from the familiar, invariant PSF of Fourier optics (in that case, notions of Optical Transfer Function and Modulation Transfer Function cannot by applied in classical sense as discussed in Ref. [1]). It may also be noted that for the sake of illustration, the latter expression can be modified in order to have the PSF always centered on the optical axis of the system, whatever the real object position $\mathbf{s_O}$:

$$\text{PSF}_G(\mathbf{s},\mathbf{s_O}) = \text{PSF}_T(\mathbf{s}) \left| \sum_{n=1}^{N} a_n \exp[i\varphi_n] \exp[-ik\mathbf{s}\,\mathbf{O'P'_n}/m] \exp[ik\mathbf{s_O}(\mathbf{OP_n} - \mathbf{O'P'_n}/m)] \right|^2. \tag{3}$$

### 2.3 Maximal achievable Field of View

We may intuitively define a "maximal achievable Field of View" for the multi-aperture optical system as the image that would be formed under the following hypotheses:

- No physical diaphragm of any kind (stops, mirrors edge or central hole) is taken into account.
- The sky object is uniformly bright over a $2\pi$-steradian solid angle, hence $O(\mathbf{s_O}) = 1$.
- The optical system is of infinite size and free from aberrations, thus diffraction can be neglected and $\text{PSF}_T(\mathbf{s})$ is assumed to be the Dirac distribution $\delta(\mathbf{s})$.

Under such assumptions the maximal achievable FoV is deduced from Eq. (1), whose integral reduces to:

$$\text{FoV}(\mathbf{s}) = \left| \sum_{n=1}^{N} a_n \exp[i\varphi_n] \exp[ik\mathbf{s}(\mathbf{OP_n} - \mathbf{O'P'_n}/m)] \right|^2. \tag{4}$$

## 3 APPLICATIONS

In this section are provided several examples of applications of the here above formulas. We first review the major difference between a Fizeau (§ 3.1) and Michelson (§ 3.2) interferometer, before studying the cases of nulling telescopes (§ 3.3) and axially combined interferometers (§ 3.4). The maximal diameter of an individual telescope is assumed to be 5 m and the numerical values of F and $F_C$ are respectively equal to F = 50 m and $F_C$ = 10 mm, leading to an optical compression factor $m$ = 1/500. All other used numerical parameters are summarized in Table 1 for the various considered cases. Computations are carried out at a wavelength $\lambda$ = 10 µm, which is a typical figure for nulling interferometry in the mid-infrared band.

Table 1: Numerical values of main physical parameters for all simulated cases.

| Case | Number of entrance pupils | Number of exit pupils | B (m) | D (m) | B' (mm) | D' (mm) | Maximal FoV (arcsec)[1] | Section |
|---|---|---|---|---|---|---|---|---|
| Fizeau interferometer | 8 | 8 | 20 | 5 | 40 | 10 | Infinite | § 3.1 |
| Michelson interferometer | 2 | 2 | 20 | 5 | 20 | 10 | 0.41 | § 3.2 |
| SRT | 1 | 8 | 0 | 5 | 10 | 10 | 0.83 | § 3.3.2 |
| Sheared-Pupil Telescope | 1 | 4 | 0 | 5 | 2 | 10 | 4.13 | § 3.3.3 |
| Masked SPT | 8 | 1 | 0.5 | 4 | 0 | 8 | 8.25 | § 3.3.4 |
| Masked SPT | 8 | 1 | 1 | 3 | 0 | 6 | 4.13 | § 3.3.4 |
| ACI | 8 | 1 | 20 | 5 | 0 | 10 | 0.21 | § 3.4 |

---

1 Computed using approximate relation (9).

## 3.1 Fizeau interferometer

A generic optical scheme of stellar interferometer is shown in Figure 2, either of the Fizeau or Michelson types. Among many other definitions utilized by different authors, let us retain the following, which is illustrated in Figure 3:

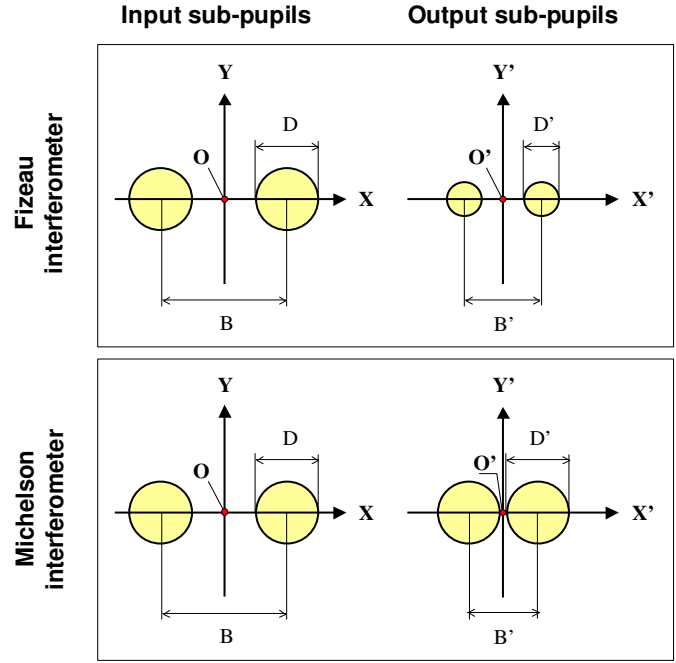

Figure 3: Input and output pupils configurations for Fizeau (top) and Michelson interferometers (bottom).

- Fizeau interferometers present the unique property that their output pupil is the scaled replica of their entrance pupil: all the entrance and exit sub-apertures as well as their relative arrangement are perfectly homothetic. Mathematically this condition implies that:

$$\mathbf{O'P'_n} = m\,\mathbf{OP_n}, \quad (1 \leq n \leq N) \tag{5}$$

also being known as the "Pupil in = Pupil out" condition [5] or "golden rule" of interferometry [6].

- In opposition the Michelson interferometer does not respect this golden rule: we then may have $\mathbf{O'P'_n} = m'\,\mathbf{OP_n}$ with any value of $m' = B'/B$ differing from $m$, or even no homothetic relationship between entrance and exit apertures at all (e.g. exit sub-apertures are often aligned along a single axis even in the cases of non-linear interferometric arrays). Alternatively, one may talk about a "pupil densification" process. The historical Michelson's 20 feet interferometer on Mount Wilson was perhaps the first example of densified optical system.

In the case of Fizeau interferometers, inserting Eq. (5) into Eq. (1) readily leads to the familiar convolution product of Fourier optics between the object O(**s**) and its image I(**s**) formed by the multi-aperture instrument:

$$I(\mathbf{s}) = \left[PSF_T(\mathbf{s})\,F(\mathbf{s})\right] * O(\mathbf{s}), \tag{6a}$$

where function:

$$F(\mathbf{s}) = \left|\sum_{n=1}^{N} a_n \exp[i\varphi_n] \exp[i\,k\,\mathbf{s}\,\mathbf{OP_n}]\right|^2 \tag{6b}$$

is named "Far-field Fringe Function" (FFF) in Ref. [1] because it describes the interference pattern that would be observed in the image plane if all sub-pupils were reduced to a pinhole[1]. It follows from Eq. (6a) that Fizeau interferometers possess the natural ability to form real images of an observed sky object, being eventually disturbed by shifted replicas of the same object generated by the FFF. Now inserting condition (5) into Eqs. (3) and (4) allows retrieving two basic properties of Fizeau interferometers:

1) The generalized PSF of the optical system $PSF_G(\mathbf{s},\mathbf{s_O})$ does not depend any longer on vector $\mathbf{s_O}$, hence the PSF is invariant over the whole interferometer Field of View.

2) The maximal achievable FoV of the Fizeau interferometer can be expressed as the very simple relationship:

$$FoV(\mathbf{s}) = \left| \sum_{n=1}^{N} a_n \exp[i\varphi_n] \right|^2, \qquad (7)$$

implying that the FoV is uniform, taking a constant value that only depends on the amplitude transmission factors $a_n$ and phase-shifts $\varphi_n$. For a classical imaging interferometer being perfectly cophased ($\varphi_n = 0$. whatever is n), $FoV(\mathbf{s})$ is uniformly equal to 1, which is in agreement with the golden rule of interferometry [5-6].

---
[1] It may be noticed that this FFF definition is somewhat related to the "dirty map" used by Högbom into his CLEAN algorithm [10].

However this golden rule may be of dramatic consequence when nulling, Fizeau-type interferometers are considered, since the phase-shifts $\varphi_n$ must be chosen so as to cancel the light originating from the central star: this has the consequence that all PSFs are invariant and equal to zero at their theoretical centers, hence any bright or faint punctual object (including planets) located anywhere in the FoV should be nulled as well as the parent star, and therefore not detectable (in that case, the principle of energy conservation suggests that most of the optical power will be reflected to the metrology sensors represented in Figure 2). To illustrate this point, let us consider the case of an eight-telescopes Fizeau interferometer in both imaging and nulling modes, following the squared arrangement sketched on the left panel of Figure 4 and using the numerical parameters of Table 1. It can be verified that the PSFs are effectively invariant in the whole achievable FoV, and that the latter is uniformly bright in imaging mode and dark in nulling mode. It may then be concluded that searching for extra-solar planets requires a deliberate violation of the famous "Pupil in = Pupil out" condition.

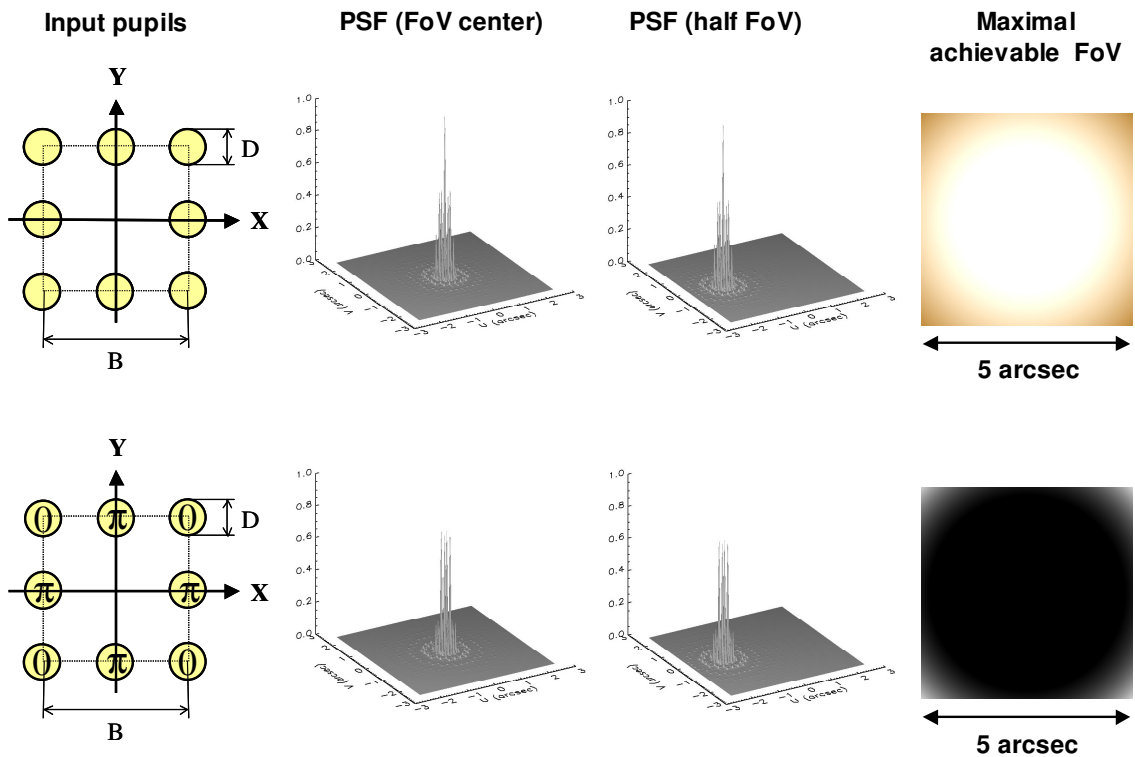

Figure 4: Invariant PSFs and achievable FoV of an eight-telescopes Fizeau interferometer in both imaging (first row) and nulling versions (bottom row). Achromatic phase-shifts $\varphi_n$ for the nulling version are indicated on left bottom panel.

## 3.2 Michelson interferometer

As mentioned in the previous subsection, the essential characteristic of a Michelson interferometer is that the golden rule is no longer being respected. If the overall geometry of the telescope array is preserved, one may write however:

$$\mathbf{O'P'_n} = m' \, \mathbf{OP_n}, \quad \text{with:} \quad m' = B'/B \neq m, \quad (1 \leq n \leq N) \quad (8)$$

The inequality between $m$ and $m'$ has a known consequence on the maximal achievable FoV that will here be considered as a sanity check of the whole presented formalism. According to Tallon-Bosc and Tallon [11] and using the here employed notations, the maximal polychromatic FoV of a Michelson interferometer is approximately:

$$\text{FoV} \leq \frac{\lambda^2}{B \, \delta\lambda} \frac{1}{(1 - m'/m)}, \quad (9)$$

where $\delta\lambda$ is the effective spectral bandwidth and B the maximal baseline between any couple of telescopes. It must be noticed that as expected this maximal FoV becomes infinite (at least theoretically) for a Fizeau interferometer where by definition $m' = m$. In the case of co-axial recombination ($m' = 0$, see § 3.4), Eq. (9) is also in agreement with Haniff's FoV definition being equal to the product of the spatial and spectral resolutions $\lambda^2 / B \, \delta\lambda$ [12]. But it is also possible to evaluate the polychromatic FoV by direct numerical integration of Eq. (4), i.e.:

$$\text{FoV}_{\delta\lambda}(\mathbf{s}) = \int_{\delta\lambda} \text{FoV}_\lambda(\mathbf{s}) \, B_{\delta\lambda}(\lambda) \, d\lambda \Big/ \int_{\delta\lambda} B_{\delta\lambda}(\lambda) \, d\lambda, \quad (10)$$

with function $B_{\delta\lambda}(\lambda)$ standing for the energetic profile of the selected spectral band[1]. Now considering a simple two-telescopes Michelson interferometer in imaging and nulling configurations, whose numerical parameters are provided in the second row of Table 1 and a spectral bandwidth $\delta\lambda = 5$ µm, we can see in Figure 5 a good agreement between the theoretical and numerical evaluations of the maximal achievable FoV. The major interest of Eq. (10) is that it can be accurately and quickly evaluated whatever the total number N of telescopes and geometrical arrangement and phase-shifts of their entrance and exit pupils.

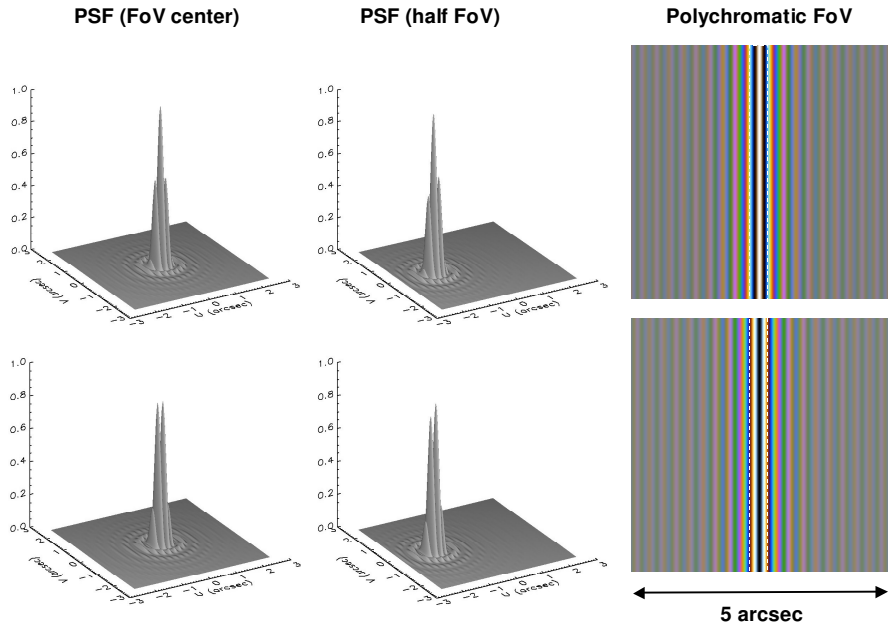

Figure 5: Variable PSFs and achievable FoV of a two-telescopes Michelson interferometer in imaging (first row) and nulling modes (bottom row). Dashed lines on the right panels indicate FoV limits.

---

[1] The same kind of numerical integration could also be performed on Eqs (2-3), allowing one to evaluate polychromatic PSFs varying in the whole interferometer FoV.

### 3.3 Nulling telescopes

Several designs of "nulling telescopes" have been proposed by different authors in the past years [8] [13-14]. The basic idea is to adapt some of the best planned technologies for sparse apertures nulling interferometry to the case of a single space telescope of moderate diameter, in order to characterize the exo-zodiacal clouds of a number of target extra-solar systems as well as their Jupiter-like planets, on the one hand, and to validate the envisaged data reduction and pseudo-imaging processes, on the other hand. Below are reviewed some of these concepts in light of the presented formalism.

#### 3.3.1 Monolithic nulling telescope

A very simple idea to turn an afocal, monolithic telescope into a nulling instrument could be to direct the output compressed beam to a Mach-Zehnder Interferometer (MZI) as depicted in the left panel of Figure 6, where a couple of Achromatic Phase Shifters (APS) made of wedged dispersive plates are introducing an achromatic, half-wave OPD difference between both interferometer arms ($\varphi_n - \varphi_{n'} = \pi$). This technique should fail, however, when the telescope input pupil (P) is optically conjugated with the MZI output pupil (P'), since the whole system finally obeys to the golden rule of Fizeau interferometers, and the destructive interference pattern should then spread out over the whole FoV as discussed in § 3.1. Hence alternative concepts have to be found.

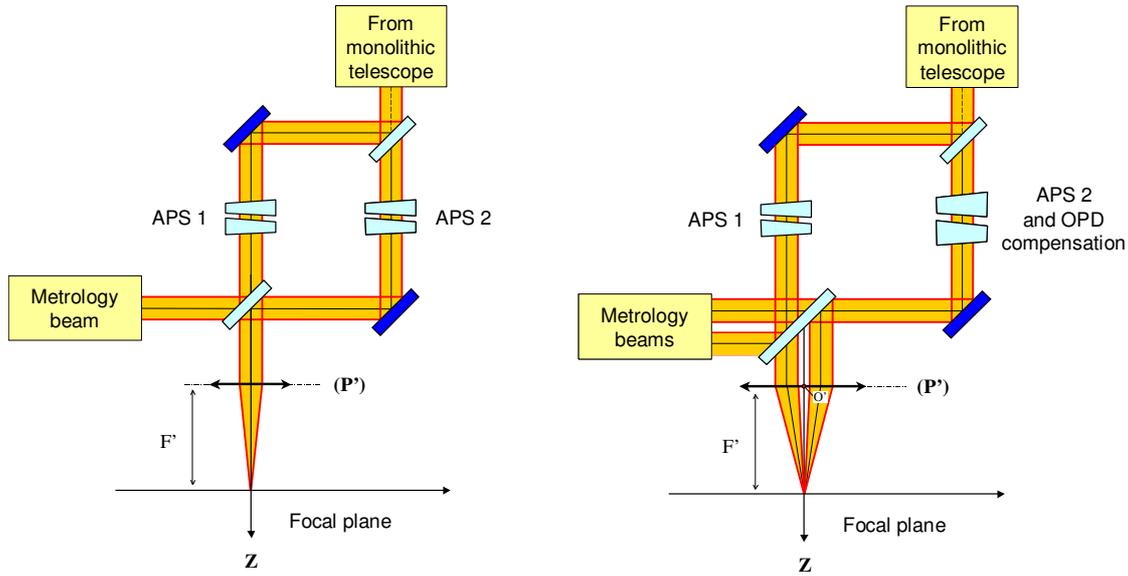

Figure 6: Recombination schemes of a monolithic nulling telescope (left) and of a Sheared-Pupil Telescope (right).

#### 3.3.2 Super-Resolving Telescope (SRT)

A tentative concept of "super-resolving telescope" (SRT) with multiple exit apertures has been previously described in Refs. [1] and [14]. It consists in one single afocal collecting telescope, optically feeding a number N of off-axis, parallel exit beams by means of cascaded beamsplitters. The beams are further expanded and multi-axially recombined in the image plane, thus enabling PSF cores to be much thinner than those generated by the monolithic telescope itself. Although this system actually overcomes Rayleigh's diffraction limit and Abbe's sine law, it has been demonstrated that it nevertheless does not provide any real super-resolution capacity, since its basic Object-Image relationship writes [1]:

$$I(\mathbf{s}) = F(\mathbf{s}) \left[ PSF_T(\mathbf{s}) * O(\mathbf{s}) \right], \text{ with far-field fringe function: } F(\mathbf{s}) = \left| \sum_{n=1}^{N} a_n \exp[i\varphi_n] \exp[-i k \mathbf{s} \mathbf{O'P'_n} / m] \right|^2. \quad (11)$$

Hence all high spatial frequency information of the object O(**s**) has already been filtered out by the monolithic telescope aperture *before* the incident optical power is finally concentrated on the thinned PSF cores. This is illustrated by the numerical simulations of Figure 8 showing typical PSF and FoV characteristics of a SRT, computed from relationships (2), (4) and (10) with the numerical parameters provided in Table 1 (the distribution of phase-shifts $\varphi_n$ in the exit aperture plane is the same as depicted in the bottom left panel of Figure 4). It can be noted that the PSFs are composed of a group of sharp diffraction lobes confined into a circular angular area corresponding to the telescope PSF, and that the peaks amplitudes are varying over the whole FoV. The monochromatic FoV itself appears as an infinite regular grid of thin bright spots, which reduces to the four central ones on a spectral bandwidth $\delta\lambda/\lambda = 50\%$. We conclude that the worst drawback of the SRT probably consists in such an extremely restricted FoV.

*3.3.3 Sheared-Pupil Telescope (SPT)*

The Sheared-Pupil Telescope (SPT) concept is indeed another example of densified optical system. It may basically be understood as a variant of the monolithic nulling telescope presented in section 3.3.1, where the output arms of the MZI are laterally shifted one with respect to the other[1] as shown on the right panel of Figure 6. Generalizing this principle to any telescope number then generates a set of N sheared sub-apertures in the output pupil plane that can interfere together (see Figure 7). Here two main types of SPT can be distinguished, depending on the presence of a Lyot stop in the output pupil plane (P'): current section deals with the case of an "unmasked SPT" (no Lyot stop in the exit pupil plane) while the case of a "masked SPT" (including the Lyot stop) is considered in section 3.3.4.

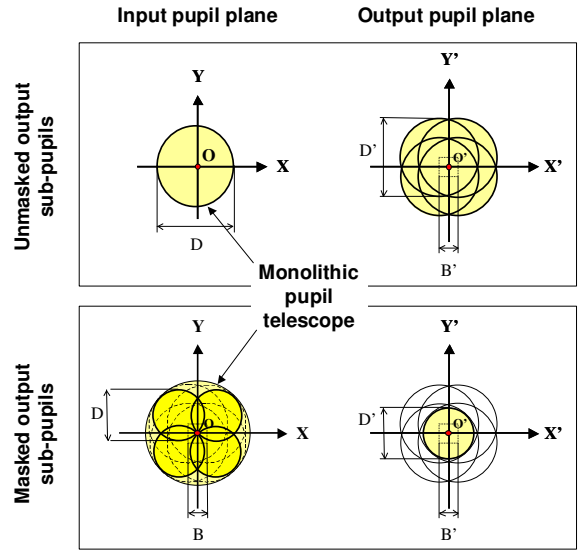

Figure 7: Two possible entrance and exit pupil arrangements for the Sheared-Pupil Telescope (N = 4).

PSF and FoV behaviors for the unmasked SPT are illustrated on the second row of Figure 8 in the case of a nulling "Angel cross" configuration (N = 4, $\varphi_1 = 0$, $\varphi_2 = \pi$, $\varphi_3 = 0$, $\varphi_4 = \pi$). Here again the PSF is varying over the whole FoV, and a noticeable star leakage appears at the FoV centre. The accessible monochromatic and polychromatic FoVs now look much more extended than those provided by the densified SRT. It must be highlighted, however, that the both systems are closely related, since they are governed by the same fundamental Object-Image relationship that is Eq. (11).

*3.3.4 Masked SPT*

With respect to the previous design, the masked SPT incorporates an additional Lyot stop placed at the exit pupil plane (P') and vignetting each sub-pupil so that they are all reduced to a common circular area of diameter D' as depicted on the bottom panel of Figure 7. This has the effect of selecting N delimited, sheared sub-areas of diameter D on the entrance pupil of the monolithic telescope (usually its primary mirror) and making them interfere together in the image plane. For this reason the imaging properties of masked SPTs will be similar to those of the Axially Combined Interferometer (ACI) presented in the next section. Imaging properties of both masked SPT and ACI concepts are further addressed in section 4.

---

[1] This lateral shift can be introduced using other interferometer types, e.g. a Michelson equipped with cube-corners as in Ref. [8].

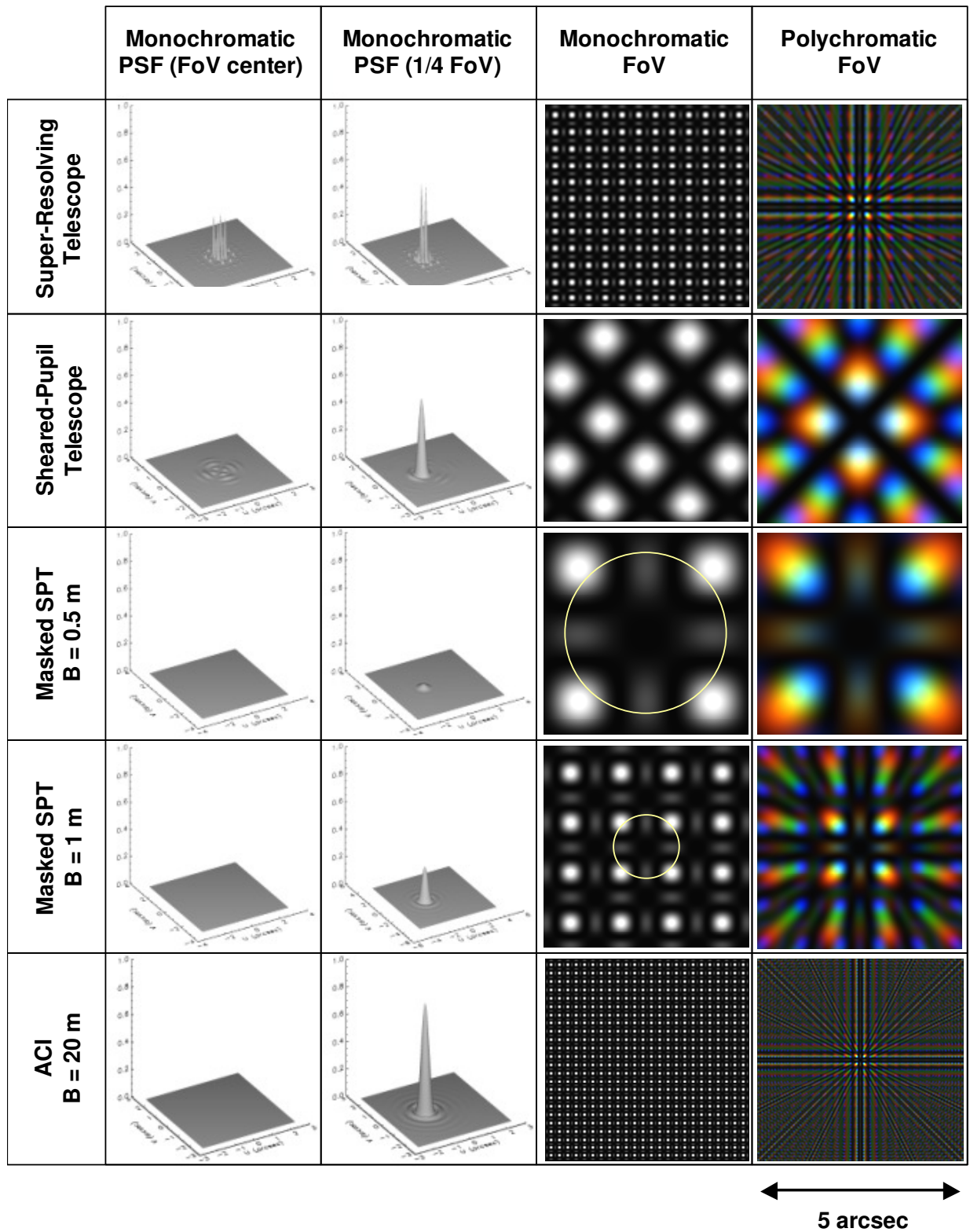

Figure 8: PSF and field of view characteristics of super-resolving telescope (first row), sheared-pupil telescopes (middle rows) and the axially combined interferometer (bottom row).

The third and fourth rows in Figure 8 display the variable PSFs and achievable monochromatic and polychromatic FoVs of a nulling, masked SPT constituted of eight entrance sup-pupils with the same phase-shifts as for the SRT case. The two major differences with respect to the unmasked SPT are as follows:

- No stellar leakage is apparent at the FoV centre, which is a much more favorable condition to reach a deep nulling of the central star.
- Unfortunately, the latter advantage is somewhat counterbalanced by the extended area of the nulled FoV where any object (including planets) will suffer from a low throughput. As an example, yellow circles in Figure 8 are materializing inner regions where throughput is always inferior to 0.25, corresponding to "darkened areas" of 4 and 1.5 arcsec for baselines B equal to 0.5 and 1 m respectively. Only the second case seems acceptable for planet searching, however it implies that the useful section of the collecting telescope is reduced to 3 m (see Table 1). Hence good sky coverage seems only attainable at the price of some losses in radiometric efficiency.

One may thus expect masked SPTs to be preferred in terms of rejection rate, and unmasked SPTs for what concerns radiometric performance (in turn improving signal-to-noise ratio and integration time). The point is further discussed in section 4, after having examined one last, classical multi-aperture optical system that is the Axially Combined Interferometer (ACI).

### 3.4 Axially Combined Interferometer (ACI)

The axially combined interferometer may be considered as a special case of Michelson interferometer where all output sub-pupils are merged together (i.e. $m' = 0$ following the herein presented formalism). Given an interferometer constituted of N separated telescopes, this condition can be realized by means of an arrangement of beamsplitters such as represented in Figure 9. A number of ACIs has already been designed and built for imaging purpose [15-16], and they can readily be turned into nulling instruments by means of an APS device as in Bracewell's original concept [7]. The specific Object-Image relationship of the ACI has been demonstrated and extensively discussed in Ref. [1], writing:

$$I(\mathbf{s}) = PSF_T(\mathbf{s}) * [F(\mathbf{s}) O(\mathbf{s})], \text{ with far-field fringe function: } F(\mathbf{s}) = \left| \sum_{n=1}^{N} a_n \exp[i\varphi_n] \exp[i\, k\, \mathbf{s}\, \mathbf{OP_n}] \right|^2 \quad (12)$$

It must be emphasized that due to the presence of the Lyot stop, Eq. (12) is also fully applicable to the masked SPT, hence both types of system share the same imaging properties that are further discussed in section 4. PSF and field of view characteristics of a nulling ACI are illustrated on the bottom row of Figure 8, showing total extinction at the FoV centre, high throughput on the constructive interference peaks, and a very thin polychromatic angular FoV varying as the inverse of the telescope baseline B as predicted by relation (9).

## 4 IMAGING PROPERTIES OF NULLING SPT AND ACI

In this section are finally provided some preliminary and qualitative interpretations of the imaging properties of nulling SPTs and ACIs. It has been mentioned in sections 3.3 and 3.4 that Object-Image relationships applicable to the unmasked SPT, on the one hand, and to both masked SPTs and ACIs, on the other hand, are defined by Eqs. (11) and (12) respectively. The latter relationship is very remarkable, since it implies that the observed sky-object is masked by the FFF of the interferometric array *before* diffraction from the single pupil of the telescope occurs. This is perhaps the fundamental reason why the masked SPT and ACI designs are so appropriate for nulling, because they allow in principle to cancel all the light originating from a bright central star regardless of diffraction effects. Another important consequence of Eq. (12) is that deep nulling should be feasible even with imperfect optics (i.e. $PSF_T(\mathbf{s})$ differing from an ideal Airy distribution), provided that the defects of all telescopes and relay optics are identical along the N interferometer arms. In practice however, this condition should still require the use of spatial or modal wavefront error filtering devices in the image plane of the interferometer, as is foreseen for all current projects.

The imaging properties of nulling SPTs and ACIs are illustrated in Figure 10 for various types and configurations summarized in Table 1. We consider successively the unmasked SPT, a masked SPT with entrance baselines B = 0.5 m and B = 1 m, and a nulling ACI with B = 10 m and B = 20 m. All computations are performed with λ = 10 µm and a maximal diameter of the individual telescopes equal to 5-m. For each case the monochromatic FoV of the instrument is displayed as a gray-scale map (yellow circle indicating central areas where planet throughput remains lower than 0.25) and 3D view. Rightmost columns reproduce images formed by all considered systems from a fake sky object whose brightness distribution is shown on the left bottom panel of Figure 10 (here we aim at nulling the central area and isolating the outermost bright star). From nulling SPTs down to ACIs, the following trends have been be noted:

1) The nulling unmasked SPT presents the advantages of a high throughput and best concentration of energy originating from the off-axis star, however there remains a residual leakage at the 5 % level from central objects.

2) The masked SPT provides full extinction of the central area, but its radiometric efficiency is found to be very low (< 1 %) for small baselines (B = 0.5 m), which confirms the conclusions of § 3.3.4. Longer baselines should be preferred despite of the decreased sub-pupils area and loss in radiometric efficiency.

3) The nulling ACI behaves as a masked SPT with high throughput advantage for very short baselines (e.g. B = 10 m where the individual telescope pupils are connected side by side). In this case the nulled FoV is significantly narrower than the SPT FoV and the central star has been fully canceled, but a 40 % leakage from the central ring is present. For much longer baselines (B > 20 m), the FoV becomes so small that the nulling capacity of the instrument is definitively lost and the direct image is the same as would be observed with an individual, monolithic telescope of 5-m diameter.

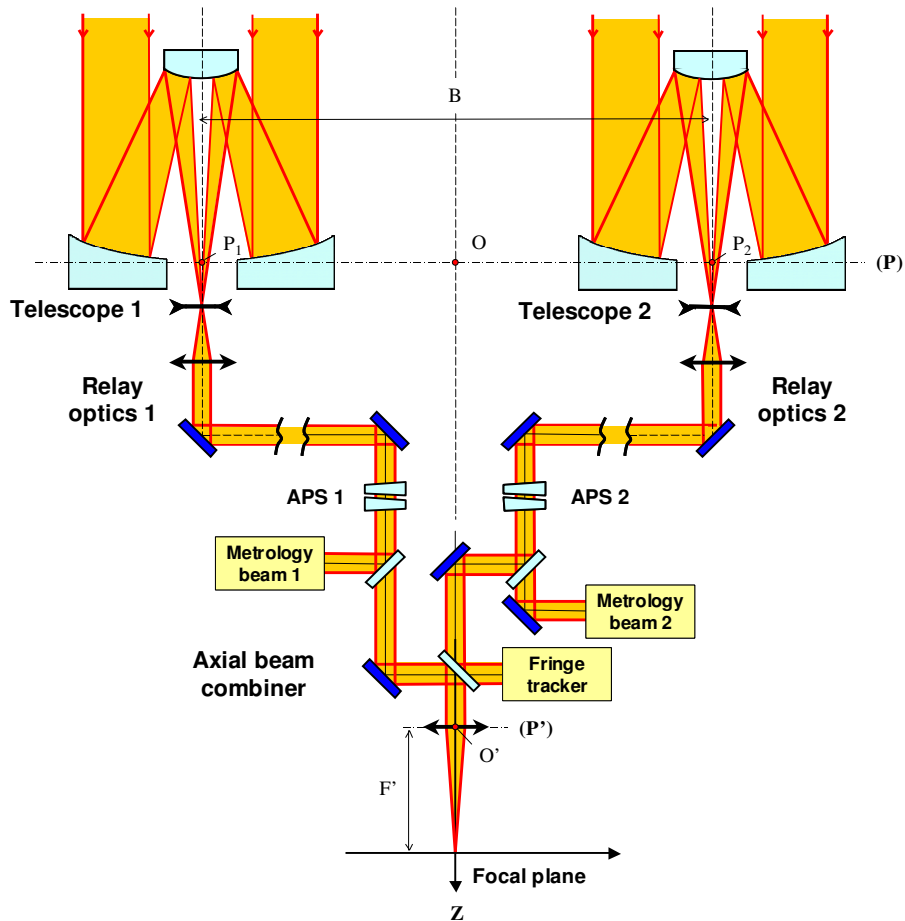

Figure 9: Schematic optical layout of an axially combined interferometer.

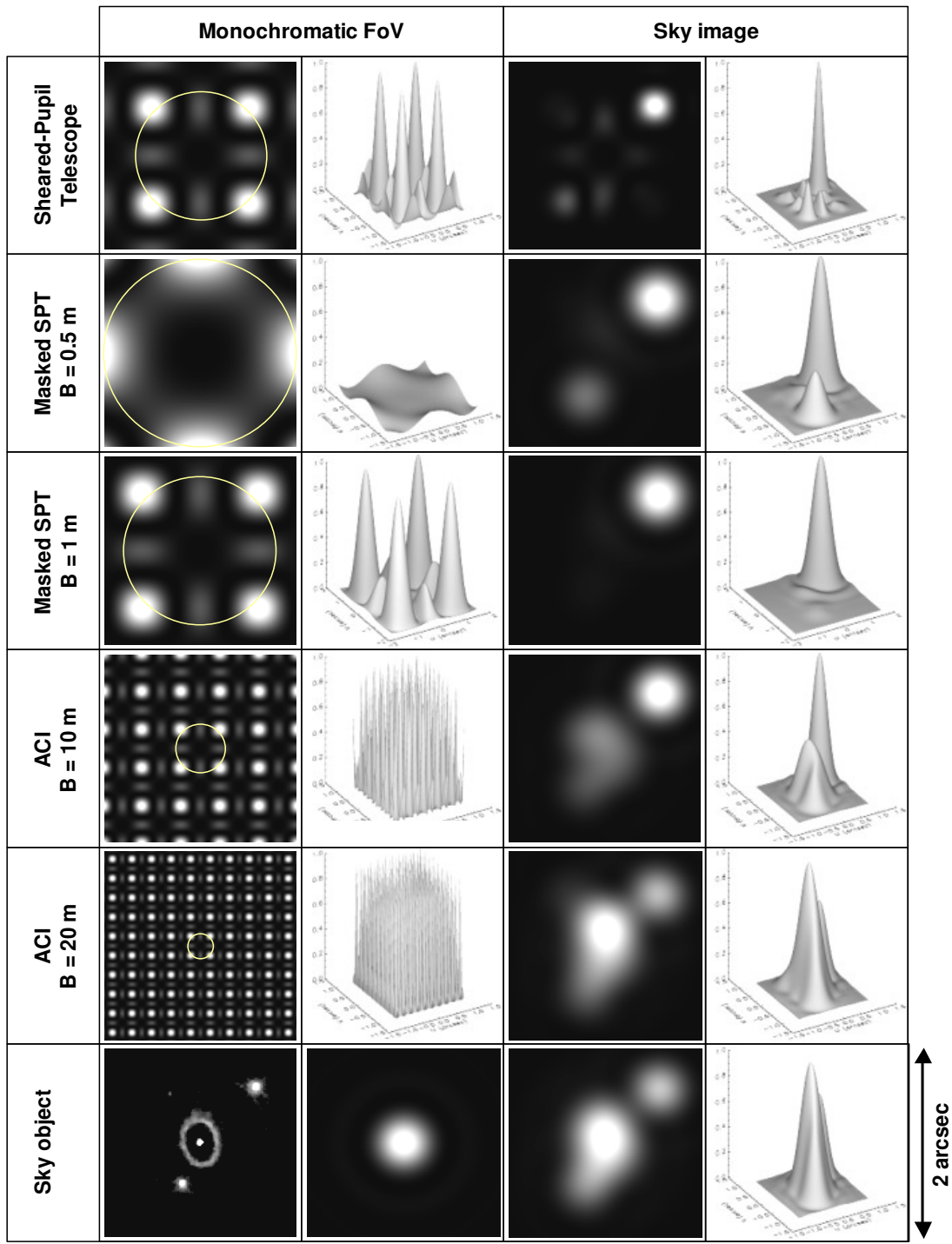

Figure 10: Imaging properties of nulling SPTs and ACIs. First row: nulling, unmasked SPT. Second and third rows: nulling masked SPT with B = 0.5 m and B = 1 m. Fourth and fifth rows: nulling ACI with B = 10 m and B = 20 m. Left columns: gray-scale maps and 3D views of monochromatic FoV. Right columns: gray-scale maps and 3D views of image projected back on-sky. Last row is showing from left to right: original sky object, PSF of an individual telescope, and image of the sky object seen through the individual telescope (gray-scale maps and 3D view). Note that the latter have been normalized to unity regardless of the actual throughput.

In view of the previous results, it might be expected that the best designs would either be a masked SPT with reduced entrance sub-pupils or an unmasked SPT associated with a robust leakage calibration procedure, depending whether deep nulling or high radiometric efficiency is to be favored. However a rigorous tradeoff between both concepts should obviously involve much more criteria such as those summarized in section 6 of Ref. [1], and is clearly beyond the scope of the present communication.

# 5 CONCLUSION

In this communication were reviewed some classical concepts of multi-aperture, imaging or nulling interferometers in the perspective of a first-order Fourier optics formalism that has been recently presented in Ref. [1]. Various topics such as fundamental difference between Fizeau and Michelson interferometers (the "golden rule" of interferometry), maximal achievable Field of View in monochromatic and polychromatic light, or imaging capacities of nulling, monolithic telescopes and axially combined interferometers were revisited. Object-Image relationships applicable to all the presented optical systems have been provided, discussed and illustrated with the help of numerical simulations. It must be emphasized that the main goal of this paper is not to select the most promising optical architecture, but simply to provide the reader with a set of quick computing tools, allowing fast and accurate calculation of the point-spread functions, field of view, and simulated images formed by these complex high angular resolution systems.